\documentclass[aps,pra,twocolumn,showpacs,floatfix]{revtex4}
\usepackage{epsfig}
\usepackage{graphicx}
\usepackage{dcolumn}
\usepackage{amsthm,amsmath}

\begin{document}

\title{An Optimized Ion Trap Geometry to Measure Quadrupole Shifts of $^{171}$Yb$^+$ Clocks}

\author{N. Batra$^{1,2}$, B. K. Sahoo$^3$, and S. De$^1$}
\affiliation{$^1$CSIR-National Physical Laboratory, Dr. K. S. Krishnan Marg, New Delhi-110012, India \\
$^2$Academy of Scientific and Innovative Research, CSIR road, Taramani, Chennai-600113, India\\
$^3$Theoretical Physics Division, Physical Research Laboratory, Navrangpura, Ahmedabad-380009, India}

\date{Received date; Accepted date}

\begin{abstract}
We propose a new ion-trap geometry to carry out accurate
measurements of the quadrupole shifts in the $^{171}$Yb-ion. This
trap will produce nearly ideal harmonic potential where the
quadrupole shifts due to the anharmonic components can be reduced
by four orders of magnitude. This will be useful to reduce the
uncertainties in the clock frequency measurements of the $6s
~{^2}S_{1/2} \rightarrow 4f^{13} 6s^2 ~{^2}F_{7/2}$ and $6s
~{^2}S_{1/2} \rightarrow 5d ~{^2}D_{3/2}$ transitions, from which
we can deduce precise values of the quadrupole moments ($\Theta$s)
of the $4f^{13} 6s^2 ~{^2}F_{7/2}$ and $5d ~{^2}D_{3/2}$ states.
Moreover, it may be able to affirm validity of the measured
$\Theta$ value of the $4f^{13} 6s^2 ~{^2}F_{7/2}$ state where
three independent theoretical studies defer almost by one order in
magnitude from the measurement. We also perform calculations of
$\Theta$s using the relativistic coupled-cluster (RCC) method. We
use these $\Theta$ values to estimate quadrupole shift that can be
measured in our proposed ion trap experiment.
\end{abstract}


\maketitle

\section{Introduction}
\label{Sec:Introduction}

Advances in trapping and laser control of a single ion
\cite{ABauch_RPP_2002,Poli_Nuovo_2013,Takamoto_CRphys_2015} began
a new era for the frequency standards in the optical range, which
are aimed to achieve a fractional accuracy of $10^{-16}-10^{-18}$.
Worldwide a number of ions, such as, $^{199}$Hg$^+$
\cite{Rosenband_Science_2008}, $^{171}$Yb$^+$
\cite{Gill_IEEE_2003,Godun_PRL_2014,Tamm_PRA_2009,Huntemann_PRL_2012,
Huntemann_PRL_2014}, $^{115}$In$^+$ \cite{Wang_OC_2007},
$^{88}$Sr$^+$ \cite{Dube_PRA_2013, Barwood_PRL_2014},
$^{40}$Ca$^+$ \cite{Huang_APB_2014,Kajita_PRA_2005}, $^{27}$Al$^+$
\cite{Chou_PRL_2010} etc. have been undertaken in the experiments
to attain such promising optical frequency standards. Among them
$^{171}$Yb$^+$ is unique in the sense that it has three potential
optical transitions that can be used for clocks
\cite{Nandy_PRA_2014}. Out of these there are two narrow $6s ~
{^2}S_{1/2} (F =0, m_F = 0) \rightarrow 5d ~ {^2}D_{3/2} (F =2,
m_F = 0)$ \cite{Webster_IEEE_2010, Tamm_PRA_2014a}, $6s ~
{^2}S_{1/2} (F =0, m_F = 0) \rightarrow 5d ~ {^2}D_{5/2} (F =2,
m_F = 0) $ \cite{Roberts_PRA_1999} quadrupole (E2) transitions and
an ultra-narrow $6s ~ {^2} S_{1/2} (F =0, m_F = 0)  \rightarrow
4f^{13} 6s^2 ~ {^2}F_{7/2} (F =3, m_F= 0)$ octupole (E3)
transition \cite{King_NJP_2012,Huntemann_PRL_2014} with their
respective wavelengths at 435.5 nm, 411 nm and 467 nm. The
transitions at the wavelengths 435.5 nm and 467 nm with low
systematic shifts are the most suitable ones for precision
frequency standards owing to their extremely small natural
line-widths 3.02 Hz and 1 nHz, respectively. Their precisely
measured transition frequencies $\nu_o$ have already been reported
as $688\, 358\, 979 \, 309\, 306.62 (73)$ Hz \cite{Tamm_PRA_2014a}
and $688\, 358\, 979 \, 309\, 310 (9)$ Hz \cite{Webster_IEEE_2010}
for the E2-transition and $642\, 121\, 496\, 772\, 645.36 (39)$ Hz
\cite{Huntemann_PRL_2014} and $642\, 121\, 496\, 772\, 646.22
(67)$ Hz \cite{King_NJP_2012} for the E3-transition. These two
transitions are endorsed by the international committee for weight
and measures (CIPM) for secondary representation of standard
international (SI) second owing to their least sensitive to the
external electromagnetic fields. Other than being a potential
candidate for frequency standards, Yb$^+$ is also being considered
for studying parity non-conservation effect
\cite{Dzuba_PRA_2011,Sahoo_PRA_2011}, violation of the Lorenz
symmetry \cite{VADzuba_Arxiv_2015}, searching for possible
temporary variation of the fine structure constant
\cite{Dzuba_PRA_2003,Dzuba_PRA_2008} \emph{etc}.

In an atomic clock, the measured frequency of the interrogated
transition is always different than its absolute value due to the
systematics. The net frequency shift depends upon many
environmental factors and experimental conditions, which may or
may not be canceled out at the end. Thus, they need to be
accounted to establish accurate frequency standards. Careful
design of the ion traps are also very important for minimizing the
systematics caused by the environmental factors
\cite{Nisbet_Arxiv_2015}. Electric quadrupole shift is one of the
major systematics when the states associated with the clock
transition have finite quadrupole moments ($\Theta$s). In the
$^{171}$Yb$^+$ ion, the experimentally measured $\Theta$ value of
the $5d ~ {^2}D_{3/2}$ state \cite{Schneider_PRL_2005} is 50 times
larger than the $4f^{13} 6s^2 ~ {^2}F_{7/2}$ state. A recent
theoretical study suggests a more precise $\Theta$ value of the
$5d ~ {^2}D_{3/2}$ state \cite{Nandy_PRA_2014}. Similarly, three
independent theoretical investigations
\cite{Blythe_JPB_2003,Porsev_PRA_2012,Nandy_PRA_2014} shows very
large disagreements than the measured \cite{Huntemann_PRL_2012}
$\Theta$ value of the $4f^{13} 6s^2 ~{^2}F_{7/2}$ state.
Therefore, it is imperative to carry out further investigations to
attain more precise and reliable values of $\Theta$s and probe the
reason for the anomalies between the calculated and experimental
results. Here we also perform another calculation using the
relativistic coupled-cluster (RCC) method to evaluate the $\Theta$
values of the $5d ~ {^2}D_{3/2}$ and $4f^{13} 6s^2 ~ {^2}F_{7/2}$
states to use them in the present analysis. We also analyze in
detail about the suitable confining potentials, electric fields
and field gradients that can create a nearly ideal quadrupole trap
condition for carrying out precise measurements of the $\Theta$
values. Using these inputs we estimate typical values of the
quadrupole shifts of the $ 6s ~{^2}S_{1/2} (F =0, m_F = 0)
\rightarrow 5d ~ {^2}D_{3/2} (F =2, m_F = 0)$ and $6s ~ {^2}
S_{1/2} (F =0, m_F = 0) \rightarrow 4f^{13} 6s^2 ~ {^2}F_{7/2} (F
=3, m_F= 0)$ clock transitions considering a number of ion trap
geometries and discuss their possible pros and cons to make an
appropriate choice. This analysis identifies a suitable geometry
of the end cap ion trap to measure $\Theta$s of the $5d ~
{^2}D_{3/2}$ and $4f^{13} 6s^2 ~ {^2}F_{7/2}$ states of Yb$^+$
which is being developed at the National Physical Laboratory
(NPL), India \cite{ARastogi_Mapan_2015,SDe_currentsc_2014}.

%
\section{Electric Quadrupole Shift}
\label{Seec:EQS}

Electric quadrupole shift $\Delta \nu_Q$ to an atomic state with
angular momentum $F$ arises due to the interaction of the
quadrupole moment $\Theta(\gamma,F)$ with an applied external
electric field gradient $\nabla E$, where $\gamma$ represents for
the other quantum numbers of the state. A non-zero atomic angular
momentum results in a non-spherical charge distribution, thus atom
acquires higher order moments. Following this it is advantageous
to choose states with $J<1$ or $F<1$ in a clock transition for
which $\Theta= 0$. However, the excited states of the $6s ~
{^2}S_{1/2} (F =0, m_F = 0) \rightarrow 5d ~ {^2}D_{3/2} (F =2,
m_F = 0)$ and $6s ~ {^2} S_{1/2} (F =0, m_F = 0)  \rightarrow
4f^{13} 6s^2 ~ {^2}F_{7/2} (F =3, m_F= 0)$ clock transitions have
$J=3/2; F=2$ and $J=7/2; F=3$ resulting in nonzero quadrupole
shifts. These shifts can be estimated by calculating the
expectation value of the Hamiltonian given by
\cite{Ramsey_oxford_1956}
\begin{eqnarray}
H_{Q} = \mbox{\boldmath$\nabla$}E \cdot \mbox{\boldmath$\Theta$}(\gamma,F) = \sum^{2}_{q=-2}
(-1)^{q}\nabla E_{q}\Theta_{-q},
\end{eqnarray}
where ranks of the $\mbox{\boldmath$\nabla$}E$ and $\mbox{\boldmath$\Theta$}$ tensors are two and their components are indicated by subscript $q$.
The expectation value of $H_Q$ in reduced form can be expressed as \cite{Itano_Nist_2000}
\begin{eqnarray}\label{quadshiftformula}
\langle \gamma J F m_{F}|H_{Q}|\gamma J F m_{F}\rangle &=& \Theta(\gamma,J) \, \mathcal{F}_Q(I,J,F,m_{F})\nonumber\\
&~& \times \sum_{q=-2}^{2}\nabla E_{q}D_{0q},
\end{eqnarray}
where $m_{F}$ is the magnetic quantum number, $D_{0 q}$ are the
rotation matrix elements of the projecting components of $\nabla
E$ in the principal axis frame that are used to convert from the
trap axes to the lab frame \cite{Edmonds_princeton_1974},
$\Theta(\gamma,J)$ is the quadrupole moment of the atomic state
with angular momentum $J$ and
\begin{eqnarray}
\mathcal{F}_Q = (-1)^{I+J+F}(2F+1)\left(
\begin{array}{ccc}
  F & 2 & F \\
 -m_{F} & 0 & m_{F} \\
\end{array}%
\right)\nonumber\\
\times {\left(%
\begin{array}{ccc}
  J & 2 & J \\
 -J & 0 & J \\
\end{array}%
\right)}^{-1} \left\{
\begin{array}{ccc}
  J & 2 & J \\
  F & I & F \\
\end{array}\right\}.
\end{eqnarray}
Here the quantities within $( \, )$ and $\{ \, \}$ represent the
$3j$ and $6j$-coefficients, respectively. Both the excited states
of the above mentioned clock transitions acquire
$\mathcal{F}_Q=1$. Due to axial symmetry of the trap, the
frequency shift contributions from $D_{0 \pm 1}$ cancel with each
other, thus finite contributions comes only from the $D_{00} = (3
\cos^2\theta -1)/2$ and $D_{0 \pm 2} = \sqrt{3/8}\sin^2\theta
(\cos 2\phi \mp i \sin 2\phi)$ components, for the Euler's angles
$\theta$ and $\phi$. In this work, we use our calculated $\Theta$
values for the $5d ^{2}D_{3/2} (F = 2)$ and $4f^{13}
6s^{2}~^{2}F_{7/2} (F = 3)$ states to find out the optimum
electrode geometries that can produce nearly-ideal quadrupole
confining potentials after interacting with the resultant electric
field gradients of the non-ideal multipole potentials $\Phi(x,y,z)
= \sum_{k=1}^{\infty} \Phi^{(k)}$ with the order of multipole $k$
of the effective trapping potentials.

%
\section{Methods for calculation}
\label{Sec:Theory}

To calculate atomic state wave functions for the determination of the $\Theta(\gamma,J)$ values we adopt the Bloch's approach \cite{lindgren}.
Following this approach we express the wave function $\vert \Psi_v \rangle$ of the $5d ~{^2}D_{3/2}$ state with the valence orbital $v$ in
the $5d_{3/2}$ orbital as
\begin{eqnarray}
 \vert \Psi_v \rangle = \Omega_v \vert \Phi_v \rangle
\end{eqnarray}
and the wave function $\vert \Psi_a \rangle$ of the $4f^{13}6s^2~{^2}F_{7/2}$ state with the valence orbital $a$ in the $4f_{7/2}$
orbital as
\begin{eqnarray}
 \vert \Psi_a \rangle = \Omega_a \vert \Phi_a \rangle
\end{eqnarray}
of the Yb$^+$ ion, where $\Omega_v$ and $\Omega_a$ are the wave
operators for the corresponding reference states $\vert \Phi_v
\rangle$ and $\vert \Phi_a \rangle$, respectively. We use two
different ways to construct these reference states. For the
computational simplification we choose the working reference
states as the Dirac-Hartree-Fock (DHF) wave functions of the
closed-shell configurations (denoted by $\vert \Phi_0^{v/a}
\rangle$) in place of the above mentioned respective actual
reference states $\vert \Phi_v \rangle$ and $\vert \Phi_a \rangle$
having open valence orbitals. In our calculations we have obtained
$\vert \Phi_0^{v} \rangle$ for the $[4f^{14}]$ configuration,
while $\vert \Phi_0^{a} \rangle$ is calculated with the
$[4f^{14}]6s^2$ configuration. Then, the actual reference states
are obtained by appending the valence orbital $v=5d_{3/2}$ and
removing the spin partner of the valence orbital $a=4f_{7/2}$ of
the respective references. In the second quantization formalism,
it is given as
\begin{eqnarray}
 \vert \Phi_v \rangle =a_v^{\dagger} \vert \Phi_0^{v} \rangle \ \ \ \text{and} \ \ \ \vert \Phi_a \rangle =a_a \vert \Phi_0^{a} \rangle .
\end{eqnarray}
We employ the Dirac-Coulomb Hamiltonian for the calculations which in the atomic unit (a.u.) is given by
\begin{eqnarray}
H &=& \sum_i \left [ c\mbox{\boldmath$\alpha$}_i\cdot \textbf{p}_i+(\beta_i -1)c^2 + V_n(r_i) + \sum_{j>i} \frac{1}{r_{ij}} \right ], \ \ \ \ \ \
\end{eqnarray}
with $\mbox{\boldmath$\alpha$}$ and $\beta$ are the usual Dirac matrices and $V_n(r)$ represents the nuclear potential.

In a perturbative procedure, $\Omega_{v/a}$ can be expressed as
\begin{eqnarray}
 \Omega_{v/a} =  1+ \chi_0^{v/a} + \chi_{v/a} \equiv \Omega_0^{v/a} + \chi_{v/a},
\end{eqnarray}
where $\chi_0^{v/a}$ and $\chi_{v/a}$ are responsible for carrying out excitations from $\vert \Phi_0^{v/a} \rangle$ due to the
residual interaction $V_r=H-H_0$ for the DHF Hamiltonian $H_0$. In a series expansion they are given as
\begin{eqnarray}
 \chi_0^{v/a} = \sum_k \chi_0^{v/a(k)} \ \ \text{and} \ \ \chi_{v/a}=\sum_k \chi_{v/a}^{(k)},
\end{eqnarray}
where the superscript $k$ refers to the number of times $V_r$ is considered in the many-body perturbation theory (MBPT(k) method).
The k$^{\rm{th}}$ order amplitudes for the $\chi_0^{v/a}$ and $\chi_{v/a}$ operators are obtained by solving the following
equations \cite{lindgren}
\begin{eqnarray}
 [\chi_0^{v/a(k)},H_0]P_0^{v/a} &=& Q_0^{v/a} V_r(1+ \chi_0^{v/a(k-1)} )P_0^{v/a} \ \ \ \ \
\end{eqnarray}
and
\begin{eqnarray}
[\chi_{v/a}^{(k)},H_0]P_{v/a} &=& Q_{v/a} V_r (1+ \chi_0^{v/a(k-1)} \nonumber \\
&+& \chi_{v/a}^{(k-1)}) P_{v/a} - \sum_{m=1}^{k-1}\chi_{v/a}^{(k-m)} P_{v/a}  V_r \nonumber \\
&\times& (1+\chi_0^{v/a(m-1)}+\chi_{v/a}^{(m-1)})P_{v/a} ,
 \label{mbsv}
\end{eqnarray}
where the projection operators are defined as $P_0^{v/a}=\vert \Phi_0^{v/a} \rangle \langle \Phi_0^{v/a} \vert $, $Q_0^{v/a}= 1-
P_0^{v/a}$, $P_{v/a}=\vert \Phi_{v/a} \rangle \langle \Phi_{v/a} \vert $ and $Q_{v/a}= 1- P_{v/a}$. The exact energies for the
states having the closed-shell and open-shell configurations are evaluated using the effective Hamiltonians given by
\begin{eqnarray}
 H_0^{eff,v/a}= P_0^{v/a} H \Omega_0^{v/a} P_0^{v/a}
 \label{efhmc}
\end{eqnarray}
and
\begin{eqnarray}
 H_{v/a}^{eff}= P_{v/a} H\Omega_{v/a} P_{v/a}.
 \label{efhmv}
\end{eqnarray}

\begin{table}
\caption{Demonstration of trends in the $\Theta$ values (in a.u.)
from lower to higher order methods. Our calculations are compared
with the values available from other calculations and the
experimental results. The estimated final values are shown in bold
fonts. Uncertainties to these values are also quoted separately.}
\begin{ruledtabular}
\begin{tabular}{lcc}
 Method       & $4f^{14}5d \, \rm{^5D_{3/2}}$   & $4f^{13}6s^2 \, \rm{^2F_{7/2}}$   \\
 \hline
  & & \\
 DHF          & 2.504                           & $-0.258$                             \\
 MBPT(2)      & 2.049                           & $-0.344$                             \\
 LCCSD        & 2.028                           & $-0.230$                             \\
 CCSD$^{(2)}$ & 2.060                           & $-0.208$                             \\
 CCSD         & 2.061                           & $-0.223$                             \\
 CCSD$_{\rm{ex}}$  & {\bf 2.079}                & ${\bf -0.224}$                      \\
 & & \\
 Uncertainty  & $\pm 0.008$                     & $\pm 0.010$                          \\
 \hline
 Others       & 2.174$^a$                       & $-0.22^b$                             \\
              & 2.157$^c$                       & $-0.20^d$                             \\
              & 2.068(12)$^e$                   & $-0.216(20)^e$                        \\
 Experiment   &  2.08(11)$^f$                   &  $-0.041(5)^g$                        \\
\end{tabular}
\end{ruledtabular}
\label{Tab:tab1}
\begin{tabular}{ll}
References: & $^a$\cite{Itano_PRA_2006}. \\
            & $^b$\cite{Blythe_JPB_2003}.\\
            & $^c$\cite{Latha_PRA_2007}.\\
            & $^d$\cite{Porsev_PRA_2012}.\\
            & $^e$\cite{Nandy_PRA_2014}.\\
            & $^f$\cite{Schneider_PRL_2005}.\\
            & $^g$\cite{Huntemann_PRL_2012}.\\
\end{tabular}
\end{table}

In the RCC theory framework, the wave functions of the considered states are expressed as (e.g. see \cite{Nandy_PRA_2014,Nandy_PRA_2013})
\begin{eqnarray}
|\Psi_v \rangle =e^{T^{v}} \{ 1+S_v \} |\Phi_v \rangle \label{eq1}
\end{eqnarray}
and
\begin{eqnarray}
|\Psi_a \rangle = e^{T^{a}} \{ 1+R_a \} |\Phi_a \rangle,
\label{eq2}
\end{eqnarray}
where $T^{v}$ and $T^{a}$ excite the core electrons from the new
reference states $|\Phi_0^v \rangle$ and $|\Phi_0^a \rangle$,
respectively, to account for the electron correlation effects. The
$S_v$ and $\left ( e^{T^v} - 1 \right ) S_v$ operators excite
electrons from the valence and valence with core orbitals from
$\vert \Phi_v \rangle$. Similarly, the $R_a$ and $\left ( e^{T^a}
- 1 \right ) R_a$ operators excite electrons from the valence and
valence with core orbitals from $\vert \Phi_a \rangle$.  In this
work we have considered only the singles and doubles excitations
in the RCC theory (CCSD method), which are identified by the RCC
operators with the subscripts $1$ and $2$, respectively, as
\begin{eqnarray}
 T^{v/a} &=& T_1^{v/a} +T_2^{v/a} \nonumber \\
 S_v &=& S_{1v} + S_{2v} \nonumber \\
 \text{and} \ \ \  R_a &=& R_{1a} + R_{2a}.
\end{eqnarray}
When only the linear terms are retained in Eqs. (\ref{eq1}) and
(\ref{eq2}) with the singles and doubles excitations approximation
in the RCC theory, we refer it to as LCCSD method. The amplitudes
of the above operators are evaluated by the equations
\begin{eqnarray}
 \langle \Phi_0^{v/a,*} \vert \overline{H}_N^{v/a}  \vert \Phi_0^{v/a} \rangle &=& 0
\label{eqt} \\
 \langle \Phi_v^* \vert \big ( \overline{H}_N^{v} - \Delta E_v \big ) S_v \vert \Phi_v \rangle &=&  - \langle \Phi_v^* \vert \overline{H}_N^v \vert \Phi_v \rangle
\label{eqsv}
 \end{eqnarray}
 and
\begin{eqnarray}
 \langle \Phi_a^* \vert \big ( \overline{H}_N^a - \Delta E_a \big ) R_a \vert \Phi_a \rangle &=&  - \langle \Phi_a^* \vert \overline{H}_N^a \vert \Phi_a \rangle ,
\label{eqsv1}
 \end{eqnarray}
where $\vert \Phi_0^{v/a} \rangle$ and $\vert \Phi_{v/a}^*\rangle$ are the excited state configurations  with respect to the DHF states
$\vert \Phi_0^{v/a} \rangle$ and $\vert \Phi_{v/a}\rangle$, respectively, and $\overline{H}_N^{v/a}= \big ( H_N e^{T^{v/a}} \big )_c$ with
subscript $c$ representing for the connected terms only. Here $\Delta E_{v} = H_{v}^{eff} -H_0^{eff,v}$ and $\Delta E_{a} = H_{a}^{eff} -
H_0^{eff,a}$ are the attachment energy of the electron in the valence orbital $v$ and ionization potential of the electron in the orbital $a$,
respectively. Following Eqs. (\ref{efhmc}) and (\ref{efhmv}), $\Delta E_{v/a}$ are evaluated as
\begin{eqnarray}
 \Delta E_{v}  = \langle \Phi_{v} \vert \overline{H}_N^{v} \left \{ 1+S_v \right \} \vert \Phi_v \rangle
 \label{eqeng1}
\end{eqnarray}
and
\begin{eqnarray}
 \Delta E_{a}  = \langle \Phi_{a} \vert \overline{H}_N^{a} \left \{ 1+ R_a \right \} \vert \Phi_a \rangle .
 \label{eqeng2}
\end{eqnarray}
To improve quality of the wave functions, we use experimental values of $\Delta E_{v/a}$ instead of the calculated values in the
CCSD method and refer to the approach as CCSD$_{\text{ex}}$ method. This is obviously better than the approach to improve the
$E_{v/a}$ values by incorporating contributions from the important triple excitations in a perturbative approach in the CCSD method
(CCSD(T) method) that was employed in \cite{Nandy_PRA_2014,Nandy_PRA_2013}.

\begin{figure}[t]
\begin{center}
{\includegraphics[width=8.5cm, height=4.0cm]{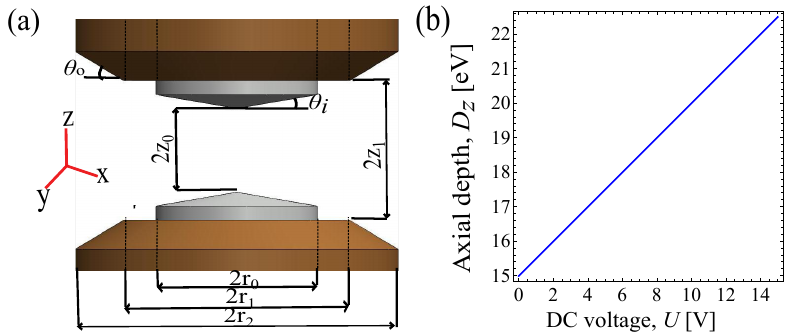}}
\end{center}
\caption{(a) Our proposed electrode assembly of the end-cap trap
with $2z_0 = 0.7$ mm, $2z_2 \approx 1.0$ mm, $2r_1 = 1$ mm, $2r_2
= 1.4$ mm, $2r_3 = 2$ mm, $\theta_i = 10^\circ$ and
$\theta_{o}=45^\circ$. (b) Variation of the trap depth as a
function of dc voltage $U$ for a fixed radio-frequency
$\omega_{rf}$ and an ac voltage $V$.}\label{Fig:figure1}
\end{figure}

After obtaining amplitudes of the MBPT and RCC operators using the equations described above, the $\Theta$ values of the considered
states are evaluated using the expression
\begin{eqnarray}
\frac{\langle \Psi_{v/a} \vert \Theta \vert \Psi_{v/a}
\rangle}{\langle \Psi_{v/a} \vert \Psi_{v/a} \rangle } &=& \frac
{\langle \Phi_{v/a} \vert \Omega_{v/a}^{\dagger} \Theta
\Omega_{v/a} \vert \Phi_{v/a} \rangle} {\langle \Phi_{v/a} \vert
\Omega_{v/a}^{\dagger} \Omega_{v/a} \vert \Phi_{v/a} \rangle } .
\label{preq}
\end{eqnarray}
This gives rise to a finite number of terms for the MBPT(2) and
LCCSD methods, but it involves two non-terminating series in the
numerator and denominator, which are $e^{T^{v/a \dagger}} \Theta
e^{T^{v/a}}$ and $e^{T^{v/a \dagger}} e^{T^{v/a}}$ respectively,
in the CCSD method. We account contributions from these
non-truncative series by adopting iterative procedures as
described in our previous works \cite{Sahoo_PRA_2015,
Singh_PRA_2015}. We also give results considering only the linear
terms of Eq. (\ref{preq}) that appear exactly in the LCCSD method,
but using amplitudes of the RCC operators from the CCSD method
(refer it as CCSD$^{(2)}$ method). Therefore from the differences
in the results between the LCCSD and CCSD methods, one can infer
about the importance of the non-linear terms in the calculations
of the wave functions; while from the differences in the results
between the CCSD$^{(2)}$ and CCSD methods, it will signify the
roles of the non-linear effects appearing in Eq. (\ref{preq}) for
the estimations of the $\Theta$ values.

We present $\Theta$ values of the $4f^{14}5d \, \rm{^5D_{3/2}}$
and $4f^{13}6s^2 \, \rm{^2F_{7/2}}$ states of Yb$^+$ in Table
\ref{Tab:tab1} from various methods as described above. We also
compare our results with the other calculations and available
experimental values. We consider results from the
CCSD$_{\text{ex}}$ method as our recommended values as this method
accounts more physical effects. We have also estimated
uncertainties to the CCSD$_{\text{ex}}$ results by estimating
neglected contributions due to truncation in the basis functions
and from the omitted correlation effects mainly that could arise
through the triply excited configurations. We had also presented
these values using the CCSD method in our previous work
\cite{Nandy_PRA_2014}, however considering only the important
non-linear terms from the $e^{T^{v/a \dagger}} \Theta e^{T^{v/a}}$
and $e^{T^{v/a \dagger}} e^{T^{v/a}}$ non-truncative series of Eq.
(\ref{preq}). As said before these series are solved iteratively
to include infinity numbers of terms in this work. Again, we have
removed uncertainties due to the calculated energies that enter
into the amplitude solving Eq. (\ref{eqsv}) of the CCSD method by
using the experimental energies.
%

%
\section{Ion Trap Induced Shift}
\label{Sec:Experiment}
\begin{figure}
\begin{center}{
\includegraphics[width=3 in]{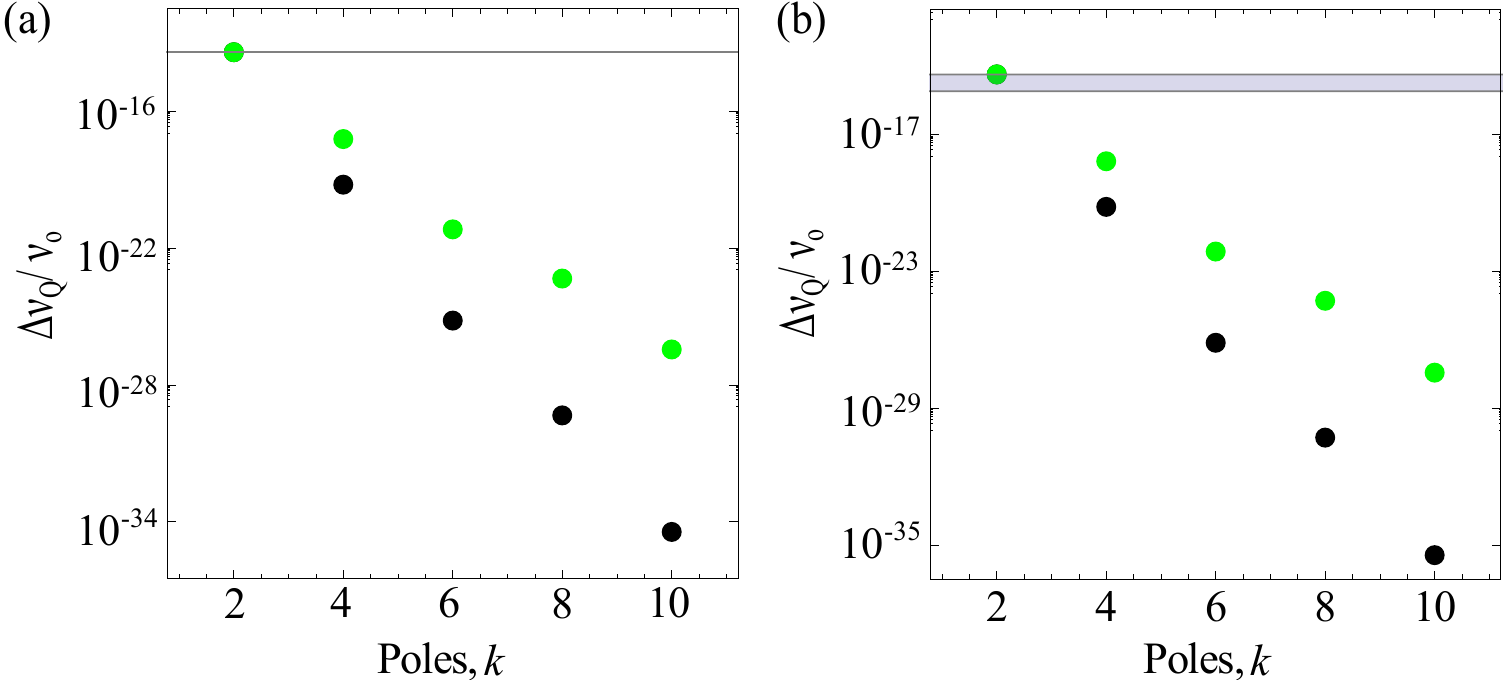}
}\end{center} \caption{(Color online) Fractional electric
quadrupole shifts due to the multipoles up to $k=10$ for the (a)
E2 and (b) E3 transitions, when the ion is off centered by a
distance 10 $\mu$m (black) and 100 $\mu$m (green), respectively.
The harmonic potential gives a spatially independent
$|\Delta\nu_Q/\nu_o|$. The reported fractional accuracies $5
\times10^{-16}$ for the E2-transition \cite{Tamm_PRA_2009}, $7.1
\times10^{-17}$ \cite{Huntemann_PRL_2012} and $1 \times10^{-15}$
\cite{King_NJP_2012} for the E3-transition are indicated by a gray
line in (a) and by a gray band in (b),
respectively.}\label{Fig:figure3}
\end{figure}

We plan to employ a modified Paul trap \cite{Paul_Rev_1990} of end
cap geometry as shown in Fig.\ref{Fig:figure1}(a). In reality,
such traps are not capable of producing pure quadrupole potential
$\Phi^{(2)}$ due to geometric modifications of the hyperbolic
electrode, machining inaccuracies and misalignments. On the other
hand, precision measurements with ions stored in a non-ideal trap,
the anharmonic components of the potential $\Phi^{(k>2)}(x,y,z)$
are non-negligible due to the fact that they change ion dynamics
and also affect the systematics. For minimizing such effects
several groups, such as, NPL UK \cite{Schrama_Optcomm_1993}, NRC
Canada \cite{Dube_PRA_2013} and PTB Germany
\cite{Stein_Thesis_2010, PTB_arxiv_2015} have come up with
different end cap trap designs for establishing single ion
frequency standards. Here we aim to identify a new end cap trap
geometry in which the trap induced quadrupole shift can be
minimized. This trap can also add minimum anharmonicity to the
confining potential and small micromotions \cite{Our_calculation}.
In a cylindrically symmetric trap as shown in
Fig.\ref{Fig:figure1}(a), only the even order multipoles
contribute.
Here, in order to estimate quality of the trap potentials we
consider $k$ up to 10 since amplitudes of $\Phi^{(k)}$ fall
drastically at higher $k$. The tensor components of $\nabla E$ for
each multipole potential $\Phi^{(k)}$ are opted from their
electric field components $E_{x,y,z}$. The corresponding
fractional quadrupole shift $\Delta \nu_{Q}/\nu_{0}$ at each $k$
is estimated from Eq. \eqref{quadshiftformula}. The variation of
$\Delta \nu_{Q}/\nu_{0}$ for all multipole potentials up to $k=10$
at two different distances from the trap center are estimated for
the E2 and E3-clock transitions, which are shown in Fig.
\ref{Fig:figure3}(a-b). The reported experimental values of the
quadrupole shifts for these two transitions are also depicted in
the same figure for the comparison.
The dominating perturbation of $\Phi^{(2)}$ arises from the
octupole term $\Phi^{(4)}$ which may largely affect the quadrupole
shift of the trapped ion frequency standards due to wrong choice
of the electrode geometry. As an example, we simplify the analysis
considering $\Phi(x,y,z) \simeq \Phi^{(2)}+\Phi^{(4)}$ since other
higher orders are less significant for the quadrupole shift (Fig.
\ref{Fig:figure3}). In the absence of any asymmetries the trap
potential can be written as
\begin{eqnarray}\label{quadrupole}
\Phi(x,y,z)&=&\frac{V_{T}(t)}{2{R}^{2}}\bigg[c_{2}(2z^{2}-x^{2}-y^{2})-\frac{c_{4}}{{R}^{2}}(3x^4+3y^4\nonumber\\
&~& +8z^4-24x^2z^2-24y^2z^2+6x^2y^2 )\bigg]
\end{eqnarray}
where, $R = \sqrt{{{r_{0}}^2}/2+{z_{0}}^2}$ and $V_{T} (t) = U + V
\cos(\omega_{rf}t)$ that depends on the dc and rf components of
the trapping voltages with amplitudes $U$ and $V$, respectively.
The dimensionless coefficients $c_{2}$ and $c_{4}$ depend on the
electrode geometry. Here we consider $|x| = |y| = r$, since the
trap is axially symmetric. The harmonic part of the potential can
produce the restoring force on the ion and the resultant axial
trap depth yields $D_z(U,V,\omega_{rf}) = U/2 + m{z_{o}}^2
{\omega}^2_{rf} {q_{z}}^2/16 Q$ \cite{Werth_springer_2010} where
$q_z = -16QV c_2/m R^2 {\omega}^2_{rf}$ and $Q$ and $m$ are the
charge and mass of the ion, respectively. As an example, in Fig.
\ref{Fig:figure1} (b) we show the variation of $D_z$ with $U$ for
fixed values $V = 500$ V and $\omega_{rf} = 2 \pi \times 12$ MHz,
respectively. Since the first order quadrupole shift from the rf
averages to zero and its second order is also zero for
$\rm{^{171}Yb^+}$ \cite{Schneider_PRL_2005}, we estimate the shift
considering $U = 10$ V. The quadrupole shift as given by $\sum_{q}
\nabla E_{q}D_{0q}$ results to $2 V_T c_2 [D_{00} -D_{02}/\sqrt{6}
]$ due to the harmonic part of the potential and it is constant
within the trapping volume. The spatial dependency comes from the
higher orders, for example $\Phi^{(4)}$ results to a quadrupole
shift of $12 V_T c_4 [4z^2(D_{00} - D_{02}/\sqrt{6}) - r^2(2D_{00}
- \sqrt{3/2}D_{02}) ] $.
\begin{figure}[h!]
\begin{center}{
\includegraphics[width=3 in]{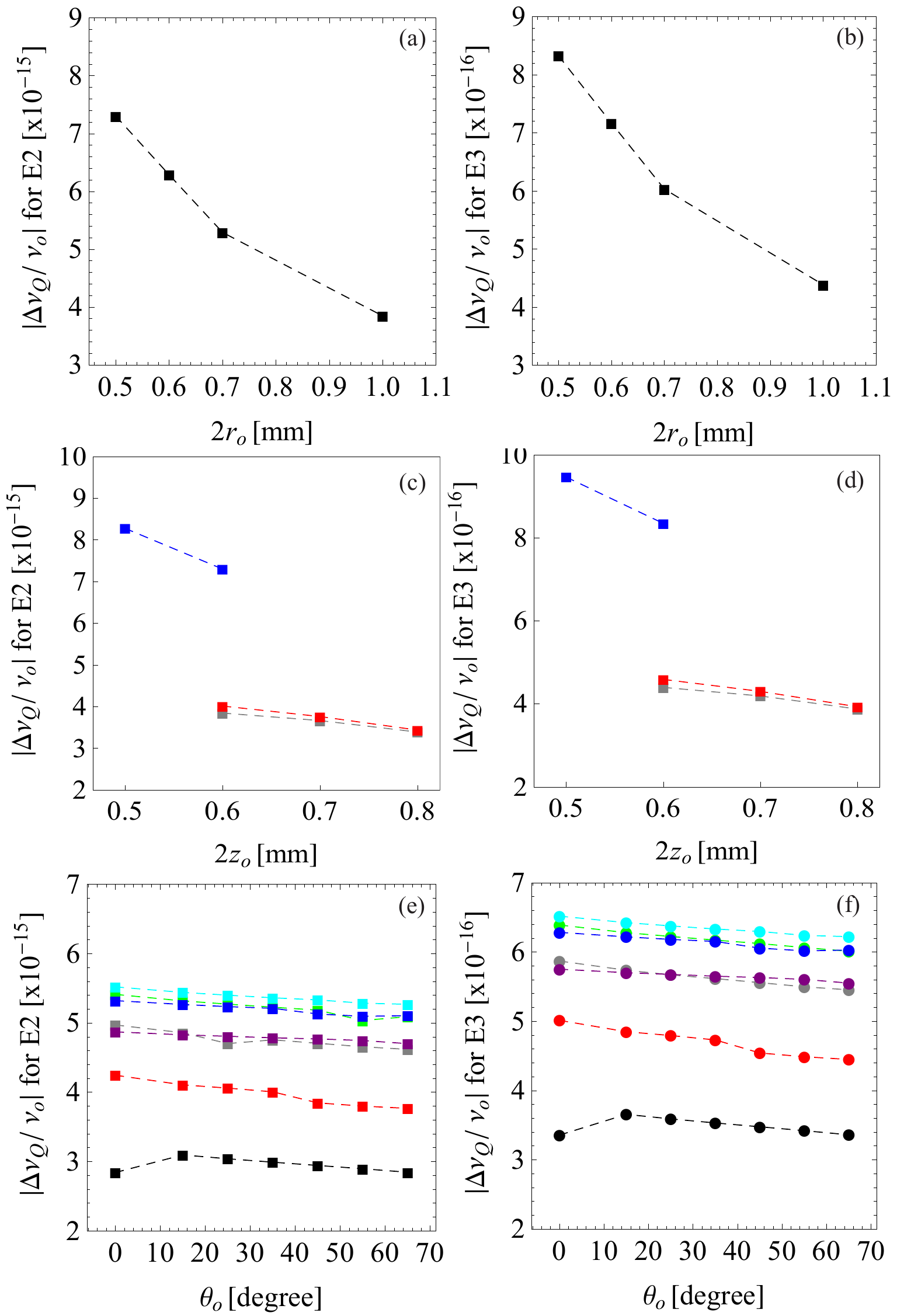}}\end{center} \caption{(Color online) Fractional electric quadrupole shift for the E2 and
E3-clock transitions corresponding to the different geometrical
parameters: (a, b) diameter $2r_o$ of the inner electrode; (c, d)
tip-to-tip separation $2z_o$ of the inner electrodes; and (e, f)
angles $\theta_o$ at which the outer electrodes are machined. Each
figure consists of a set of plots for angles of the inner
electrode $\theta_i$, which are at $0^\circ$  (black), $10^\circ$
(red), $20^\circ$ (gray), $30^\circ$ (green), $40^\circ$ (cyan),
$50^\circ$ (blue) and $60^\circ$ (purple). The dashed line
connects data points for the fixed values of $\theta_i$ and
considering $2r_o=1$ mm and $2z_o=0.7$ mm for all of them. Each
figure consists of a set of plots for the angles of the inner
electrode $\theta_i$, which are at $0^\circ$ (black), $10^\circ$
(red), $20^\circ$ (gray), $30^\circ$ (green), $40^\circ$ (cyan),
$50^\circ$ (blue) and $60^\circ$ (purple). }\label{Fig:figure2}
\end{figure}
Strength of the multipole potentials depend on $c_{k}$ as given in
Eq. \eqref{quadrupole}. The magnitudes of $c_{k}$ depend on the
geometric parameters of the trap electrodes, such as radius $r_o$,
angle $\theta_i$ of the electrode carrying rf (that is the inner
electrode), inside and outside radii $r_1$ and $r_2$, angle
$\theta_o$ of the dc carrying electrode (that is the outer
electrode) which is coaxial to the inner one and also their mutual
tip-to-tip separations $2 z_o$ and $2 z_1$, respectively, as shown
in Fig. \ref{Fig:figure1}(a). We have obtained the trap potentials
for various choices of these geometric factors by carrying out
numerical simulations using the boundary element method
\cite{CPO_USA_2013}. Then, we have characterized multipole
components in it by fitting $\sum_{k} \Phi^{(k)}$ for $k$ up to
10. Potentials are obtained for various combinations of
$\theta_i$, $\theta_o$, $2r_0$, and $2z_o$ but at the fixed values
$2r_1=1.4$ mm, $2r_2=2$ mm and $2z_1=1.16$ mm. We fix these
parameters keeping in mind that laser beams from three orthogonal
directions can impinge on the ion without any blockage as
described in Ref. \cite{ARastogi_Mapan_2015}. These three laser
beams will be used for detecting the micromotions in all the three
directions independently
\cite{Berkeland_JAP_1998,Keller_Arxiv_2015}. After studying a
series of trap geometries, we found the diameters of the outer
electrode have weak influence on $\Phi^{(k)}$ which is below the
accuracy that is expected from the machining tolerances. In Fig.
\ref{Fig:figure2}(a-d), it shows variation of the quadrupole shift
at the center of the trap with diameter and tip-to-tip separation
of the inner electrode keeping $\theta_i$ and $\theta_o$ values
fixed. Placing the inner electrodes further away from each other,
it is necessary to operate the trap at larger voltage to obtain
the required trap depth. On the other hand placing them close to
each other or having larger diameters can introduce optical
blockage for three orthogonal laser beams. Also micromotions of
the ions can increase at large $2 r_o$ and $2 z_o$
\cite{Our_calculation}. We obtain the optimized values as $2r_0 =
1$ mm and $2z_o = 0.7$ mm for which $\Delta\nu_Q$ is reduced but
not minimized. Further attempt to minimize the quadrupole shift
causes increase in anharmonicity and micromotions. Dependence of
the quadrupole shift on $\theta_i$ and $\theta_o$ are shown in
Figs. \ref{Fig:figure2} (e-f). These clearly show that the
quadrupole shift increases at large $\theta_i$ but it has
relatively weak influence on $\theta_o$. A pair of inner
electrodes with flat surfaces will introduce minimum shift.
However $\theta_i = 0$ gives optical blockage at our optimized
$2z_o = 0.7$ mm for impinging three orthogonal Gaussian laser
beams of waist $\sim 30 \, \mu$m on the ion and overlapping them
with the other laser beams. To avoid the optical blockage we have
optimized the values of $\theta_i$ and $\theta_o$ at $10^\circ$
and $45^\circ$, respectively. These would produce insignificant
number of scattered photons from the tails of Gaussian laser beams
which will propagate along the three mutually orthogonal
directions in our described design reported in Ref.
\cite{ARastogi_Mapan_2015}. For our trap geometry the coefficients
$c_2/2R^2$ and $c_4/2R^4$ are estimated to be $0.93 \times 10^6$
m$^{-2}$ and $0.11 \times 10^{12}$ m$^{-4}$, respectively.
\begin{figure}
\begin{center}{
\includegraphics[width=3 in]{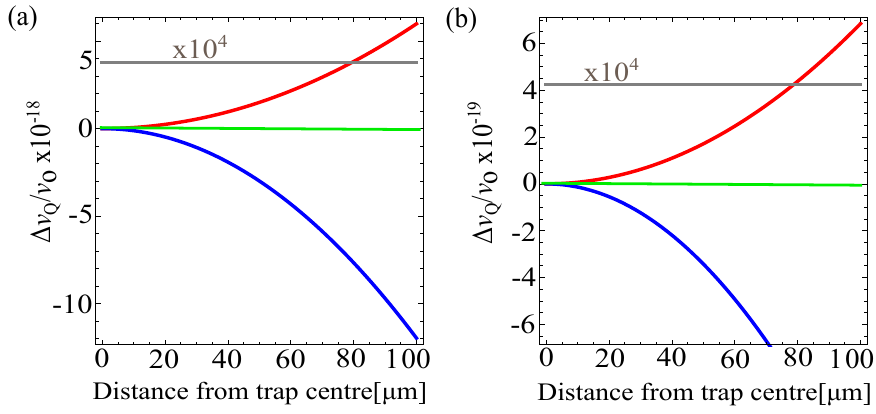}
}\end{center} \caption{(Color online) Spatial dependence of the
fractional electric quadrupole shifts $\Delta \nu_Q/\nu_o$ for the
E2 and E3-clock transitions of $\rm{^{171}Yb^+}$. The quadrupole
trapping potential produces a constant shift (gray). The spatial
dependency in the shift along the radial (red) and axial (blue)
directions arise from the anharmonic components $k>2$ of
$\Phi^{(k)}$. }\label{Fig:figure4}
\end{figure}
Due to its residual thermal motion of the ion after the laser
cooling, it is unlikely to probe the clock transition while the
ion is sitting at the center of the trap. Also there is a
possibility that the mean position of the ion is few 10 $\mu$m
away from the trap center due to imperfect stray field
compensation. This results to a spatially dependent $\Delta
\nu_{Q}$ in a non-ideal ion trap. At any position, the shift due
to potentials for $k>2$ increases following a power law of order
$k-2$ to the separation of the ion from the trap center, as shown
in Fig. \ref{Fig:figure3} (a, b) for our optimized trap geometry.
Figure \ref{Fig:figure4} shows spatial variation of $\Delta
\nu_{Q}/\nu_{0}$ due to $\Phi^{(4)}$ and compares that with shift
resulting from $\Phi^{(2)}$. The shift due to $\Phi^{(4)}$ is
found to be four orders of magnitude smaller in our particular
trap design than the contribution due to $\Phi^{(2)}$. However,
our analysis shows that a wrong choice of the trap electrode
geometry could easily increase $\Delta \nu_Q$ resulting from
$\Phi^{(4)}$ by few orders of magnitude. The quadrupole shift due
to $\Phi^{(2)}$ can be eliminated by measuring the clock frequency
along the three mutually orthogonal orientations in the lab frame
while quantizing the ion using the magnetic fields of equal
amplitudes \cite{Itano_Nist_2000,Berkeland_JAP_1998}. In Fig.
\ref{Fig:figure4}, we show such an angular averaging can eliminate
the quadrupole shift resulting from $\Phi^{(4)}$ provided the trap
assembly is perfectly axially symmetric. In practice the trap can
deviate from such an ideal situation which could lead to
inaccuracy in eliminating the quadrupole shift by angular
averaging. In such a non ideal trap the inaccuracies of
eliminating the quadrupole shift are generally induced by
$\Phi^{(4)}$. Thus the present analysis also helps us in
understanding to build a suitable trap electrode design where the
effect of $\Phi^{(4)}$ in the quadrupole shift can be reduced.
The formula for evaluating the quadrupole shift given by
\begin{eqnarray}\label{shift}
\Delta \nu_Q = \frac{c_2 U}{R^2} \times \mathcal{F}_Q \times
\Theta.
\end{eqnarray}
Substituting values of $\Theta$s from Table \ref{Tab:tab1}, we
estimate these shifts for a constant $c_2 U/R^2 = 932$ V/cm$^2$
which is expected in our trap geometry at $U=10$ V. All the
resultant quadrupole shifts for the E2 and E3-clock transitions
are shown in Figs. \ref{Fig:Qshift} (a) and (b), respectively.
This shows that in our trap we will be able to measure the
quadrupole moment of the $5d~^{2}D_{3/2} (F = 2)$ state an
accuracy of 1 part in $10^3$. This uncertainty will be one order
of magnitude better than the previous measurement
\cite{Schneider_PRL_2005}. Similarly, the quadrupole moment of the
$4f^{13} 6s^2 ~ ^{2}F_{7/2} (F = 3)$ state was previously measured
with $~12\%$ accuracy. We are also expecting to improve accuracy
of this quantity using our proposed ion trap. This will help to
verify the reported discrepancies among the experimental and
theoretical results.
\begin{figure}
\begin{center}{
\includegraphics[width=3 in]{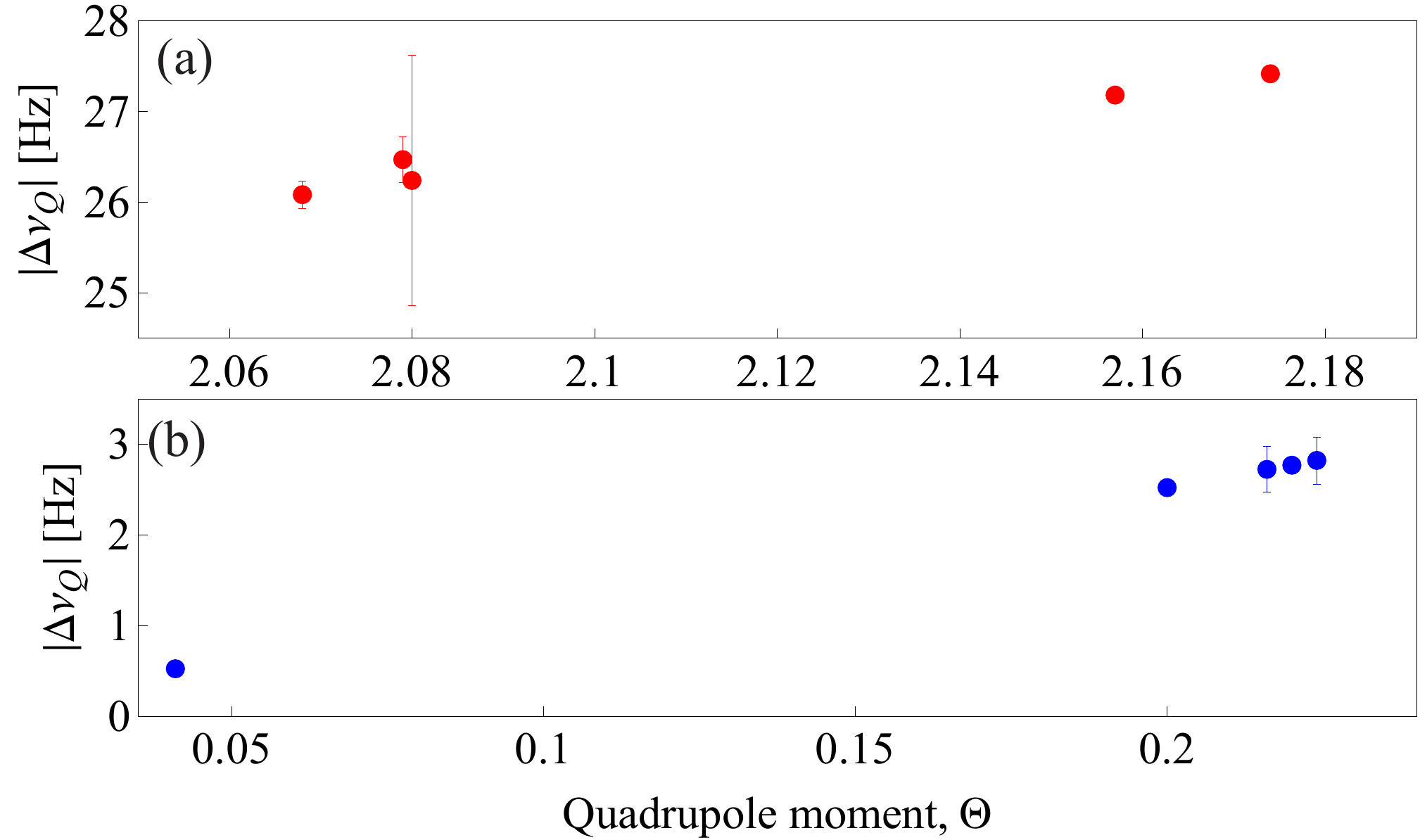}
}\end{center} \caption{(Color online) The electric quadrupole
shifts due to the previously reported works and estimated values
from this work are shown in (a) for the $6s ~ {^2}S_{1/2}
\rightarrow 5d ~ {^2}D_{3/2}$ and (b) for the $6s ~ {^2}S_{1/2}
\rightarrow 4f^{13}6s^2 ~ {^2}F_{7/2}$ clock transitions.
}\label{Fig:Qshift}
\end{figure}
%

%
\section{Conclusion}
\label{Sec:Conclusion}
We have proposed a suitable ion trap geometry for carrying out
accurate measurements of the quadrupole shifts of the $6s ~
{^2}S_{1/2} \rightarrow 5d ~ {^2}D_{3/2}$ and $6s ~ {^2}S_{1/2}
\rightarrow 4f^{13}6s^2 ~ {^2}F_{7/2}$ clock transitions in the
$^{171}$Yb$^+$. We have also carried out calculation of $\Theta$
values of the $5d ^{2}D_{3/2}$ and $4f^{13} 6s^2 ~ {^2}F_{7/2}$
states of $^{171}$Yb$^+$ using the RCC method that are used in our
analysis. We have identified an end cap ion trap geometry which
can produce nearly ideal harmonic confinement to minimize the
electric quadrupole shift. We also showed that a wrong choice of
the ion trap geometry would increase the quadrupole shift
resulting from $\Phi^{(4)}$ by two orders magnitude higher than
our optimized geometry. To get a figure of merit we have estimated
the quadrupole shifts along with their uncertainties in our
proposed setup and compared the result with the previously
available values. In our optimized ion trap we are expecting to
measure the quadrupole moment of the $5d ^{2}D_{3/2}$ state with
an accuracy one part in $10^3$. We are also aiming to measure the
quadrupole moment of the $4f^{13} 6s^2 ~ {^2}F_{7/2}$ state
reliably that could possibly explain the discrepancy between the
experimental and theoretical results.
\section*{Acknowledgement}
SD acknowledges CSIR-National Physical Laboratory, Department of
Science and Technology (grant no. SB/S2/LOP/033/2013) and Board of
Research in Nuclear Sciences (grant no. 34/14/ 19/2014-BRNS/0309)
for supporting this work. BKS acknowledges using Vikram-100 HPC
cluster of Physical Research Laboratory for the calculations.
%

%

\end{document}